# REDUNDANCY ANALYSIS AND MITIGATION FOR MACHINE LEARNING-BASED PROCESS MONITORING OF ADDITIVE MANUFACTURING

**Jiarui Xie[1] and Yaoyao Fiona Zhao[1,*]**
[1] McGill University, QC, Canada

**ABSTRACT**

*The deployment of machine learning (ML)-based process monitoring systems has significantly advanced additive manufacturing (AM) by enabling real-time defect detection, quality assessment, and process optimization. However, redundancy is a critical yet often overlooked challenge in the deployment and operation of ML-based AM process monitoring systems. Excessive redundancy leads to increased equipment costs, compromised model performance, and high computational requirements, posing barriers to industrial adoption. However, existing research lacks a unified definition of redundancy and a systematic framework for its evaluation and mitigation. This paper defines redundancy in ML-based AM process monitoring and categorizes it into sample-level, feature-level, and model-level redundancy. A comprehensive multi-level redundancy mitigation (MLRM) framework is proposed, incorporating advanced methods such as data registration, downscaling, cross-modality knowledge transfer, and model pruning to systematically reduce redundancy while improving model performance. The framework is validated through an ML-based in-situ defect detection case study for directed energy deposition (DED), demonstrating a 91% reduction in latency, a 47% decrease in error rate, and a 99.4% reduction in storage requirements. Additionally, the proposed approach lowers sensor costs and energy consumption, enabling a lightweight, cost-effective, and scalable monitoring system. By defining redundancy and introducing a structured mitigation framework, this study establishes redundancy analysis and mitigation as a key enabler of efficient ML-based process monitoring in production environments.*

Keywords: Additive manufacturing; machine learning; process monitoring; data quality; redundancy.

## 1. INTRODUCTION

Additive manufacturing (AM) has emerged as a transformative manufacturing paradigm that enables the direct fabrication of components from digital designs through a layer-by-layer material deposition process [1]. Unlike conventional subtractive manufacturing methods, AM facilitates the production of geometrically complex structures with minimal material waste and reduced tooling requirements [2]. Despite these advantages, there remain several critical challenges such as process variability, defect formation, and part quality inconsistencies, necessitating extensive post-processing and quality assurance measures [3]. Variations in thermal conditions, powder distribution, and energy input during fabrication can lead to defects such as porosity, lack of fusion, residual stresses, and surface roughness deviations [4, 5]. The complex nature of defect formation underscores the necessity for in-situ AM process monitoring and control strategies to enhance process reliability and ensure part repeatability.

To address these challenges, in-situ process monitoring has been increasingly integrated into AM systems to enable real-time modeling of process dynamics [5, 6]. A variety of sensing modalities, including optical, acoustic, and thermal sensors, are employed to capture melt pool behavior, powder distribution, and thermal gradients [7]. The data collected from these monitoring systems provide valuable insights into process stability, enabling early detection of anomalies and facilitating process optimization. Machine learning (ML) techniques have been leveraged to analyze in-situ sensor data for defect detection, anomaly classification, and process parameter optimization [8]. By establishing correlations between process conditions and part quality, ML-based approaches can evaluate and improve part quality without post-processing such as heat treatment and ex-situ metrology. The integration of ML models with process monitoring systems enables adaptive control that adjusts process parameters in real time to mitigate defects and improve consistency [9].

Despite their potential for quality assurance and enhancement, ML-based process monitoring systems remain time and cost inefficient for real-world production due to redundancies in deployment and operation [10, 11]. Figure 1 lists the aspects of deployment and operation affected by redundancy, which are further illustrated below. The most advanced monitoring systems often integrate multiple high-resolution sensors of various modalities, significantly increasing the deployment costs. However, it has been discovered that physical phenomena emitted from the same AM process can be highly correlated and thus exhibit redundant information [12-14]. During data acquisition for model training, conventional design of experiment (DOE) methods are likely to yield redundant samples by oversampling specific groups, including certain geometries, process conditions, or anomaly classes [15]. These redundant samples provide limited new information to the dataset while introducing representation bias, which can compromise the performance of ML models. Furthermore, redundancy extends to operator training, as the need to deploy multiple sensors and manage excessive data volumes increases the complexity of monitoring system deployment.

The operation of ML-based process monitoring must be both rapid, to facilitate real-time evaluation and control in dynamic manufacturing environments, and energy-efficient, to ensure cost-effectiveness and sustainability [10]. However, redundancy contributes to inefficiencies in data transmission, computation, data management, and system maintenance.

Redundant modalities and samples generate excessive data volumes, increasing the time and energy required for data transmission and handling [12]. Advanced ML models with deep architectures often incorporate an excessive number of parameters, many of which are redundant, thereby increasing prediction runtime and computational overhead [16]. Additionally, managing large datasets demands greater effort in data integration and necessitates extensive storage capacity. Maintenance costs also rise due to the redundant hardware and software components within the monitoring system. In conclusion, redundancy is the key limiting factor for ML-based process monitoring systems, as the associated costs cannot always be offset by the benefits of improved part quality. Furthermore, the increased prediction latency caused by redundant data and excessive model complexity makes it challenging for ML-based monitoring systems to effectively capture and respond to the real-time dynamics of AM processes.

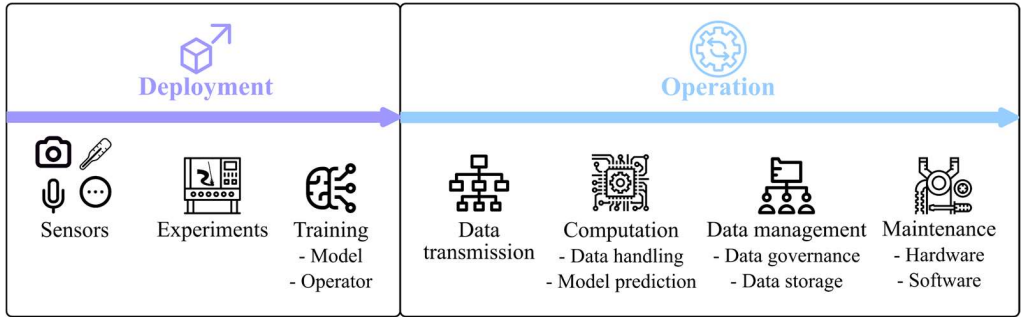

Figure 1: Aspects of ML-based process monitoring systems affected by redundancy.

Redundancy, a measure of duplicativeness, is a frequently researched concept with various definitions across disciplines such as ML, information theory, and data analytics. In ML, redundancy often refers to overlapping or unnecessary features within a dataset that do not contribute additional information, potentially leading to overfitting and reduced model performance [17]. In information theory, redundancy quantifies the extra length of a message beyond what is strictly necessary to convey the intended meaning, reflecting the extent of unnecessary representations that can be eliminated without information loss [18]. In data analytics, redundancy typically pertains to the unnecessary repetition of data within databases, which can lead to inconsistencies and inefficiencies [15, 19]. Despite extensive studies on redundancy analysis and mitigation methods for ML-based engineering process modeling [8, 15, 20, 21], a unified and comprehensive definition of redundancy in this interdisciplinary domain has yet to be established. In this context, we define redundancy as the presence of duplicative, correlated, or non-essential components in data or models, which can lead to inefficiencies in data acquisition, transmission, computation, and storage.

Analyzing and removing redundancy can significantly enhance the efficiency of ML-based process monitoring systems. In ML-based AM process monitoring, researchers have proposed various methods to mitigate sample-level [12, 20, 22], feature-level [23-25], and model-level [16, 26, 27] redundancy to achieve higher accuracy and faster prediction. However, existing studies primarily focus on individual redundancy types, proposing specialized mitigation techniques without a holistic understanding of redundancy across different levels. This paper defines redundancy in the context of ML-based process monitoring, identifies its primary sources, and proposes a framework that integrates redundancy mitigation methods. The proposed framework aims to minimize redundancy in ML-based process monitoring, significantly reducing costs and latency while improving overall system efficiency. As most studies continue to incorporate additional sensors and increasingly complex ML structures, excessive redundancy is being introduced into these systems. This paper aims to raise awareness of redundancy and establish mitigation strategies to facilitate the real-world deployment and operation of ML-based process monitoring systems. The contributions of this paper are summarized below:

- We define sample-level, feature-level, and model-level redundancy, and analyze the sources of redundancy in the context of ML-based process monitoring.
- We propose a framework that systematically reduces redundancy across all levels, leveraging advanced redundancy evaluation and mitigation methods.
- We conduct a case study to illustrate the performance improvements offered by redundancy mitigation, including reduced latency, error rate, and required storage.

The remainder of this paper is structured as follows. Section 2 reviews the literature on redundancy evaluation and mitigation methods in this domain. Section 3 identifies and defines the sources of redundancy and elaborates on the proposed redundancy evaluation and mitigation framework. Section 4 demonstrates the proposed methods in a case study of ML-based in-situ directed energy deposition (DED) defect detection. Section 5 discusses the time and cost efficiency improvements offered by redundancy mitigation. Section 6 highlights the remarks of this study.

## 2. RELATED RESEARCH

Researchers in this domain have noticed the costs, latency, and performance compromise caused by redundancy and have proposed mitigation methods at the sample, feature, and model levels. However, existing studies focused on addressing specific redundancy-induced issues, without a clear and holistic understanding of redundancy.

**Sample-level redundancy** arises when some samples introduce little new information to the dataset. These redundant samples, which cannot improve the modeling accuracy, might induce representation bias to the dataset and eventually lead to a biased model [15]. Researchers have proposed active learning and data augmentation techniques to mitigate sample-level redundancy for ML-based AM process monitoring. Active learning is a data-driven sampling method that efficiently selects the most informative data points for annotation, minimizing labeling effort while maximizing model performance. van Houtum and Vlasea [23] proposed adaptive weighted uncertainty sampling (AWUS), an active learning method that optimally balances uncertainty with random sampling to enhance model performance while minimizing annotations of redundant samples. Applied to DED quality prediction, AWUS reduces the need for labeled data by 20–70% in most experiments while outperforming six state-of-the-art sampling strategies across multiple datasets.

Data augmentation is commonly used to mitigate redundancy and representation bias when additional data acquisition or annotation is unavailable [8]. This approach synthesizes new samples for underrepresented groups to address data imbalance. According to prior research, data augmentation techniques can be classified into three categories: domain knowledge-based, statistical, and ML-based methods [15]. Domain knowledge-based methods involve manually synthesizing underrepresented samples using expert knowledge. For example, Becker et al. [28] applied signal-processing techniques such as time-stretching and amplitude-shifting to generate in-process AM acoustic signals. Similarly, Wong et al. [29] used zooming, rotation, and flipping to create synthetic AM x-ray computed tomography (XCT) images. Statistical methods either oversample underrepresented groups or undersample overrepresented groups, with techniques such as synthetic minority oversampling technique (SMOTE), statistical shape analysis (SSA), bootstrapping, and stratified sampling being widely applied in this domain for quality prediction and defect detection [8]. More recently, ML-based approaches, including generative adversarial networks (GANs) [30], autoencoders (AEs) [31], and diffusion models [32], have been implemented to generate synthetic defect images to mitigate class imbalance where defective samples are significantly fewer than defect-free ones.

**Feature-level redundancy** concerns redundancy between input features (i.e., variables) of a dataset. A large model is usually required for a multimodal dataset with high dimensionality, which leads to high risk of overfitting if not trained using a large dataset [21]. However, the input dimensionality can often be reduced without significant information loss using feature engineering, as the input features are highly-correlated in AM process monitoring datasets [20]. Feature engineering methods in this domain can be categorized into domain knowledge-based and feature learning [21, 33]. Domain knowledge-based methods derive physics-informed or statistical features based on AM, image analysis, and signal processing knowledge. For example, the geometric features of melt pools such as the area, length, and width are extracted for process anomaly detection [14, 22, 34]; Surface textual features are extracted to detect surface defects of the parts; Time and frequency domain features are derived from the raw audio signals for part quality prediction [22, 35-37]. Feature learning leverages statistical learning approaches to extract salient features based on the dataset [21]. For example, principal component analysis (PCA) techniques have been frequently utilized to derive the most representative features and reduce the input dimensionality [38-40]. ML-based feature learning algorithms such as AEs and their variants are recently employed to learn low-dimensional representations of the input data for redundancy mitigation [41, 42]. Recently, cross-modality knowledge transfer (CMKT) has been proposed to derive the features of one modality from other modalities based on the high correlation between them, thus removing the redundant modalities [12-14].

Advanced ML models employing deep architectures commonly exhibit duplicative and unnecessary elements (e.g., parameters and submodules), leading to **model-level redundancy**. In ML research, overparameterization, which describes the scenario in which the model has more parameters than needed to fit a dataset, is a frequently discussed model-level redundancy [43]. Overparametrized models are likely to overfit the training set, consume excessive computational resources, and face optimization challenges [44]. In the field of ML-based AM process monitoring, researchers have proposed model pruning and knowledge distillation techniques to address overparameterization. Zhu et al. [27] applied model pruning and operator fusion to optimize the inference time of their ML models on the edge computing board, significantly reducing the processing time from 233.01 milliseconds to 50.97 milliseconds. Shi et al. [26] proposed a knowledge distillation-based information sharing framework that transfers knowledge from data-rich units to data-poor units without direct data sharing to preserve data privacy. The results demonstrated that the proposed method reduced computational costs by 25%, as it required six times less training data while maintaining competitive accuracy, thereby improving efficiency in model training. Li et al. [16] utilized knowledge distillation to develop a compressed energy consumption prediction model for powder bed fusion by employing a teacher-student architecture. Their distilled student model achieved comparable performance with reduced computational costs, improving the time and cost efficiencies.

Despite various mitigation methods proposed in this domain, researchers still lack awareness of redundancy. While existing studies implicitly perform redundancy analysis and mitigation, none explicitly addresses it, nor recognizes it as a key barrier to ML-based process monitoring in real production environments. To bridge this gap, this paper introduces a comprehensive definition of redundancy and a multi-level redundancy mitigation (MLRM) framework, integrating previous research into a unified perspective. This study aims to establish redundancy analysis and mitigation as a fundamental enabler for efficient development of ML-based process monitoring. This methodology can be extended to various domains, such as the Internet of Things (IoT), cyber-physical systems (CPS), and digital twins (DTs), where ML-based process monitoring plays a critical role.

## 3. METHODOLOGY

This section defines redundancy and introduces its categorization within the context of ML-based process monitoring. It further discusses evaluation and mitigation methods for each redundancy type. Subsequently, an MLRM framework is proposed to systematically eliminate redundancy, enhancing both model accuracy and computational efficiency in ML-based process monitoring systems.

### 3.1. Definition and categorization of redundancy

We formulate redundancy as the addition of a component (e.g., a sample, feature, or model parameter) that does not improve modeling performance or the informational value of the dataset. Let $\mathcal{C}$ be a component added to a dataset or a model. Let $\mathcal{K}$ represent the dataset or model before adding $\mathcal{C}$. Let $P(\mathcal{K})$ be a performance metric (e.g., accuracy), a similarity metric (e.g., pairwise distance), or an information-theoretic measure (e.g., entropy) evaluated on $\mathcal{K}$. Component $\mathcal{C}$ is redundant if adding it does not significantly improve the performance or informational value of the model or dataset, in which redundancy can be expressed as:

$$R(\mathcal{C},\mathcal{K}) = 1 - \left[\left(P(\mathcal{K} \cup \mathcal{C}) - P(\mathcal{K})\right)/(|P(\mathcal{K})| + \epsilon)\right], \quad (1)$$

where $\epsilon$ is a small constant to avoid division by zero and is omitted in the following expressions for simplicity. This measure quantifies redundancy by evaluating the difference in the performance before and after adding $\mathcal{C}$, $P(\mathcal{K} \cup \mathcal{C}) - P(\mathcal{K})$, which is a novelty measure. $R(\mathcal{C},\mathcal{K})$ is interpreted as follows:

1) When $R(\mathcal{C},\mathcal{K})<1$: $\mathcal{C}$ is not fully redundant. A smaller value indicates smaller redundancy and greater performance improvement.
2) When $R(\mathcal{C},\mathcal{K})=1$: $\mathcal{C}$ is fully redundant but does not impact performance.
3) When $R(\mathcal{C},\mathcal{K})>1$: $\mathcal{C}$ is fully redundant. A larger value signifies higher redundancy leading to greater performance degradation.

Note that Equation (1) assumes that $P(\mathcal{K} \cup \mathcal{C}) - P(\mathcal{K}) > 0$ indicates improved performance. Equation (1) must be adjusted to $R(\mathcal{C},\mathcal{K}) = 1 - \left[(P(\mathcal{K}) - P(\mathcal{K} \cup \mathcal{C}))/(|P(\mathcal{K})| + \epsilon)\right]$ if $P(\mathcal{K} \cup \mathcal{C}) - P(\mathcal{K}) < 0$ indicates improved performance (e.g., mean squared error (MSE) loss). Redundancy can be a conditional or relative measure:

- **Conditional redundancy:** ML-based process modeling is usually subject to a specific modeling task (e.g., a target variable $Y$) or a condition. A component can be non-redundant for some tasks while considered redundant for other tasks. In this case, conditional redundancy is expressed as $R_{cond}(\mathcal{C},\mathcal{K},Y) = 1 - \left[(P(\mathcal{K} \cup \mathcal{C}|Y) - P(\mathcal{K}|Y))/|P(\mathcal{K}|Y)|\right]$.
- **Relative redundancy:** Biases in the dataset and model can induce imbalance between subgroups ($\mathcal{G}_1$ and $\mathcal{G}_2$), thus leading to the scenario that one subgroup is more redundant than the other: $R_{rel}(\mathcal{G}_1,\mathcal{G}_2) = P(\mathcal{G}_1)/P(\mathcal{G}_2)$.

Redundancy in this context can be classified into sample-level, feature-level, and model-level redundancy, each of which can be further subdivided based on the elements and applications of ML-based process monitoring (Figure 2). Table 1 overviews the categorization and definitions of redundancy along with the implications of redundancy mitigation.

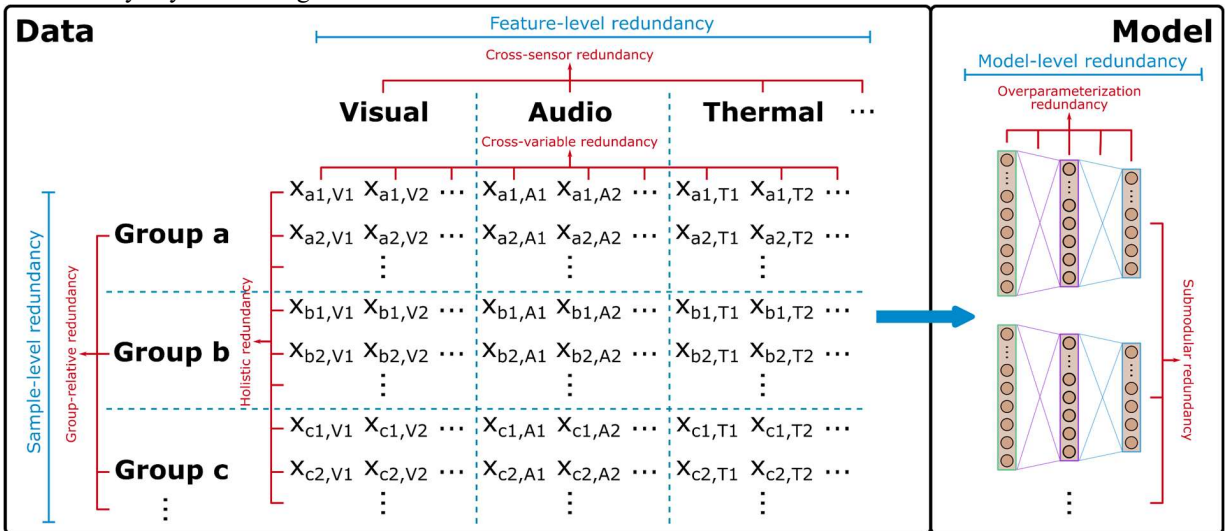

Figure 2: Categorization of redundancy in ML-based process monitoring.

Table 1: Categorization, definitions, and implications of redundancy mitigation.

| Redundancy | | Definition | Implications of mitigation |
|---|---|---|---|
| Sample-level | Holistic ($R_h$) | Redundancy of a batch of samples when added to a dataset. | Reducing representation bias and data acquisition, data storage, and computational costs. Improving model performance. |
| | Group-relative ($R_{gr}$) | Relative redundancy between data subgroups. | Reducing representation bias and improving model performance. |
| Feature-level | Cross-variable ($R_{cv}$) | Redundancy of a feature when added to a feature set. | Reducing computational costs and required data storage. Improving model performance. |
| | Cross-sensor ($R_{cs}$) | Redundancy between two feature sets from two separate sensors. | Reducing computational costs, hardware costs, and required data storage. Optimizing sensor configurations. |
| Model-level | Overparameterization ($R_{op}$) | Unnecessary parameters for fitting the training data. | Reducing computational costs and required memory space. |
| | Submodular ($R_{sm}$) | Unnecessary submodules for serving the intended function. | Reducing computational costs and required memory space. |

### 3.1.1. Sample-level redundancy

**Sample-level redundancy** refers to the presence of duplicate, highly similar, or non-informative samples in a dataset, which can increase storage requirements, computational costs, and processing time without improving model performance. Moreover, bias-induced sample-level redundancy can introduce biases into ML models, ultimately compromising their accuracy and generalizability [15]. At the sample level, redundancy can be categorized into holistic redundancy and group-relative redundancy (Figure 2).

**Holistic redundancy** ($R_h$) evaluates the redundancy of a batch of samples ($X' = \{x_1, x_2, x_3, ..., x_b\}$) with a sample size of $N_b$ when it is added to the original dataset $X_0$, which is expressed as $R_h(X', X_0) = 1 - [(P(X_0 \cup X') - P(X_0))/|P(X_0)|]$. In this context, information entropy, diversity, and coverage metrics can be utilized to compute $P(\bullet)$. Information entropy measures the uncertainty in a dataset ($X$): $\mathcal{H}(X) = -\sum_{x \in X} p(x) \log p(x)$ [18]. By analyzing how the entropy changes when a batch of samples is added or removed, we can infer whether the samples contribute new information or are redundant: $R_h(X', X_0) = 1 - [(\mathcal{H}(X_0 \cup X') - \mathcal{H}(X_0))/|\mathcal{H}(X_0)|]$. Sample-wise diversity metrics are effective means of evaluating redundancy, where higher diversity indicates lower redundancy within the dataset. For example, the similarity between two samples can be indicated using pairwise distance, $S(x_i, x_j)$, where $x_i$ and $x_j$ are two samples and $S(\bullet)$ is a distance metric such as Euclidean and Manhattan distances [15]. A simple method to evaluate the diversity of a dataset is the average pairwise distance: $\bar{S} = 2(\sum_{i=1}^{N} \sum_{j=i+1}^{N} S(x_i, x_j))/[N(N-1)]$, where N is the sample size of the dataset. Thus, diversity-based holistic redundancy is $R_h(X', X_0) = 1 - [(\bar{S}(X_0 \cup X') - \bar{S}(X_0))/|\bar{S}(X_0)|]$. Similarly, coverage metrics, which evaluate the region of the feature space covered by a dataset, can be used to determine whether the covered region expands after adding $\mathcal{C}$, thereby assessing its redundancy [15].

**Group-relative redundancy** ($R_{gr}$) is a relative measure that evaluates redundancy differences between subgroups within a dataset. In ML-based AM process monitoring, a dataset can usually be divided into subgroups in many ways, including different defect classes, materials, geometries, machines, and process conditions. Biases introduced during data acquisition may result in varying levels of redundancy across subgroups, even though holistic redundancy has been mitigated. This imbalance can compromise model generalizability, reducing its ability to perform consistently across different subgroups. Therefore, group-relative redundancy must be addressed through redundancy evaluation and mitigation to enhance model performance. An example of a simple measure to evaluate the presence of a subgroup ($\mathcal{G}_i$) is representation rate: $P(\mathcal{G}_1) = |\mathcal{G}_i|/N$, where N is the sample size of the entire dataset. The disparity ratio between two subgroups can represent the group-relative redundancy: $R_{gr}(\mathcal{G}_1, \mathcal{G}_2) = P(\mathcal{G}_1)/P(\mathcal{G}_2)$. A value greater than one indicates that $\mathcal{G}_1$ is more redundant than $\mathcal{G}_2$; therefore, $\mathcal{G}_1$ needs to be undersampled and/or $\mathcal{G}_2$ needs to be oversampled to mitigate relative redundancy, and vice versa. Similarly, performance metrics like diversity and coverage can be utilized to replace representation rate and evaluate relative redundancy.

Active learning and data augmentation are two effective approaches to mitigate sample-level redundancy and enhance model performance [15]. By incorporating redundancy measures into the adaptive sampling process, active learning can selectively acquire or annotate new samples with minimal holistic redundancy, while balancing group-relative redundancy. Redundancy measures can also guide data augmentation to mitigate representation bias by selecting the most representative subset or synthesizing novel samples with low redundancy. The data acquisition and model training costs for deploying ML-based process monitoring systems can be significantly reduced using redundancy-based DOEs and subset selection. Besides, operation costs can be reduced by analyzing and minimizing holistic redundancy. For instance, spatiotemporal redundancy is a special type of holistic redundancy in which samples collected in close spatial or temporal proximity exhibit high dependence, leading to redundant information. This is particularly relevant for sensors with high sampling rates, such as high-speed cameras and high-quality microphones. By selecting an optimal sampling rate that minimizes holistic redundancy, costs associated with excessive data acquisition, transmission, and processing can be effectively eliminated.

### 3.1.2. Feature-level redundancy

**Feature-level redundancy** occurs when multiple features in a dataset are highly correlated or non-informative, leading to increased model complexity, longer training times, reduced interpretability, and higher requirements on the dataset [21]. At the feature level, redundancy can be classified into cross-variable redundancy and cross-sensor redundancy (Figure 2).

**Cross-variable redundancy** ($R_{cv}$) investigates the redundancy within a set of features, $F = \{f_1, f_2, f_3, \ldots, f_m\}$ with a dimensionality of $N_m$. Redundancy can be evaluated between two features or between one feature and a feature set. Redundancy between two features ($R_{cv}(f_k, f_l)$, where $f_k$ and $f_l$ are two features) can be evaluated using various established methods, including Pearson's correlation, Spearman's correlation, and mutual information [21, 22]. For example, the mutual information between $f_k$ and $f_l$ is:

$$I(f_k; f_l) = \mathcal{H}(f_k) - \mathcal{H}(f_k|f_l) = \mathcal{H}(f_l) - \mathcal{H}(f_l|f_k), \quad (2)$$

where $\mathcal{H}(f_k|f_l)$ is the conditional entropy of $f_k$ given $f_l$ and $\mathcal{H}(f_l|f_k)$ is the conditional entropy of $f_l$ given $f_k$. $I(f_k; f_l)$ quantifies the reduction in uncertainty about $f_k$ given knowledge of $f_l$ and vice versa to evaluate the redundancy between them. The dimensionality of input data can be reduced by removing features with high redundancy without significant information loss. Alternatively, feature importance can be evaluated by analyzing the redundancy between an input feature and the target variable ($Y$) using $R_{cv}(f_k, Y)$. Input features that have low redundancy with the target variable have low importance and can be removed without performance degradation.

Cross-variable redundancy can be assessed between a single feature and a feature set ($R_{cv}(f_k, F_l)$, where $f_k$ is a feature and $F_l$ is a feature set) to determine whether this feature should be added or removed. Conditional mutual information quantifies the additional information $f_k$ provides about $Y$ given the feature set $F_l$: $I(f_k; Y|F_l) = \mathcal{H}(Y|F_l) - \mathcal{H}(Y|f_k, F_l)$. This metric can evaluate the redundancy of a single feature conditioned on the existing features. Model predictive performance metrics can also be utilized to measure redundancy by evaluating the improvement of model performance (e.g., F1 score) once a new feature is added to the existing feature set: $R_{cv}(f_k, F_l, Y) = 1 - [(P(F_l \cup f_k|Y) - P(F_l|Y))/|P(F_l|Y)|]$. This approach has been widely adopted in wrapper-based feature selection such as forward feature selection and backward feature elimination [21].

**Cross-sensor redundancy** ($R_{cs}$) assesses the redundancy between sets of features (e.g., $F_V$ from the visual modality and $F_A$ from the audio modality), where each set is collected from a separate sensor: $R_{cs}(F_V, F_A)$. The dataset of a multisensor monitoring system is fused from multiple sub-datasets (e.g., $D_V$ from the visual modality and $D_A$ from the audio modality) of varied data sizes and dimensionalities due to different sampling frequencies and resolutions. Cross-sensor redundancy is usually evaluated after the data sizes are aligned using techniques like data registration and downsampling [45]. Similarly, mutual information ($I(F_V; F_A)$) and conditional mutual information ($I(F_V; Y|F_A)$) can measure the redundancy of adding a new sensor with features $F_V$ to the system. Model performance metrics can also be used to evaluate the redundancy of adding $F_V$ given the target variable $Y$ and existing features $F_A$: $R_{cv}(F_V, F_A, Y) = 1 - [(P(F_A \cup F_V|Y) - P(F_A|Y))/|P(F_A|Y)|]$, where $P$ is a model performance metric. Alternatively, a model-based approach can be utilized to evaluate the correlation between $F_V$ and $F_A$ by learning a mapping function $f: F_V \to F_A$. The minimum MSE between the predicted and actual values, $MSE(f(D_V), D_A)$, serves as an indicator of redundancy: a smaller MSE suggests a stronger correlation between the two sensors, meaning they carry more redundant information. The dimensionality of the two datasets must be aligned to leverage more redundancy metrics, including distance-based metric and statistical tests.

Researchers in this domain have employed various feature engineering methods to measure and mitigate cross-variable redundancy, including feature extraction, feature selection, and feature learning [7, 8]. Reducing cross-variable redundancy decreases the data volume to process, leading to lower deployment and operational costs. A specific example is spatiotemporal redundancy, a form of cross-variable redundancy where features in close spatial or temporal proximity—such as neighboring pixels—exhibit high dependence, contributing to redundant information. Note that cross-variable spatiotemporal redundancy differs from holistic spatiotemporal redundancy as the former concerns the proximity of features and the latter concerns the proximity of samples. To address cross-variable spatiotemporal redundancy, raw data are often downscaled to a lower resolution to facilitate subsequent analysis. However, cross-sensor redundancy analysis and mitigation remains rarely explored in this domain due to its complexity. Despite being less commonly applied, it has the potential to optimize sensor configurations and even eliminate unnecessary sensors, leading to improved computational performance and significant reductions in hardware costs [46, 47].

### 3.1.3. Model-level redundancy

**Model-level redundancy** refers to the presence of unnecessary or non-essential components within a trained model, such as redundant parameters, neurons, or layers, which increase data, computational, and memory requirements without significantly improving predictive performance. At the model level, redundancy can be classified into overparameterization redundancy and submodular redundancy (Figure 2).

**Overparameterization redundancy** ($R_{op}$) occurs when the parameter set of a trained model ($\Theta_0 = \{\theta_1, \theta_2, \theta_3, \ldots, \theta_p\}$ with a size of $N_p$) contains more parameters than necessary to fit the training data. This is common in modern ML, especially in deep learning, where models often have millions or even billions of parameters. Overparameterization redundancy can be evaluated by analyzing model performance regarding the target variable ($P(\bullet|Y)$) after either introducing an extra parameter set or removing a parameter subset, both of which can be expressed as $\Theta' = \{\theta'_1, \theta'_2, \theta'_3, \ldots, \theta'_s\}$ with a size of $N_s$. In this way, overparameterization redundancy, $R_{op}(\Theta', \Theta_0, Y)$, can be expressed as $1 - [(P(\Theta_0 \cup \Theta'|Y) - P(\Theta_0|Y))/|P(\Theta_0|Y)|]$ for the former and $1 - [(P(\Theta_0|Y) - P(\Theta_0 \setminus \Theta'|Y))/|P(\Theta_0|Y)|]$ for the latter. Similarly, overparameterization redundancy can be used to determine the optimal number of neurons and layers in a deep learning model, ensuring an efficient network architecture without unnecessary complexity.

**Submodular redundancy** ($R_{sm}$) assesses the redundancy among similar submodules ($\mathcal{M} = \{\mathcal{M}_1, \mathcal{M}_2, \mathcal{M}_3, \ldots, \mathcal{M}_b\}$ with a size of $N_b$) of a model or ensemble. These submodules are typically introduced to serve specific functions in ML models. For example, ensemble learning combines multiple base models to leverage their respective strengths for enhanced model accuracy and robustness. Ensemble learning also enables uncertainty quantification, increasing the explainability and trustworthiness of ML models. Multisource and multimodal inputs often necessitate multiple encoders to accommodate variations in data types and resolutions. However, when more submodules than necessary are embraced, they can lead to submodular redundancy, increasing computational costs without contributing to performance improvements. Similarly, model performance metrics can be utilized again to evaluate submodular redundancy and determine whether a submodule $\mathcal{M}_i$ should be removed: $R_{sm}(\mathcal{M}, \mathcal{M}_i, Y) = [(P(\mathcal{M}|Y) - P(\mathcal{M} \setminus \mathcal{M}_i|Y))/|P(\mathcal{M}|Y)|]$. For two submodules ($\mathcal{M}_i$ and $\mathcal{M}_j$) with the same architecture, submodular redundancy can be measured through assessing the similarity between the parameter values: $R_{sm}(\mathcal{M}_i, \mathcal{M}_j) = \|\mathcal{M}_i - \mathcal{M}_j\|_2 = \sqrt{\sum_{k=1}^{N_p}(\theta_{i,k} - \theta_{j,k})^2}$, where $\theta_{i,k}$ and $\theta_{j,k}$ are the parameters of the submodules $\mathcal{M}_i$ and $\mathcal{M}_j$, respectively.

Model-level redundancy can be mitigated either during the initial training phase through hyperparameter optimization or during the fine-tuning phase using model pruning and knowledge distillation. Reducing model-level redundancy enhances operational efficiency by lowering computational demands and reducing inference latency. Additionally, compact models enable deployment on microcomputers, reducing hardware costs and making decentralized ML-based process monitoring systems more feasible and cost-effective.

### 3.2. Multi-level redundancy mitigation framework

This section introduces an MLRM framework that cohesively integrates the redundancy mitigation methods to enhance the performance and efficiency of ML-based process monitoring systems. Figure 3 provides a graphical representation of the framework, where the top half of each block outlines the corresponding mitigation method and the bottom half highlights its intended purpose.

The MLRM framework begins by spatiotemporally aligning multisensor data collected from the original monitoring system using data registration, ensuring a synchronized dataset with equal sample sizes across modalities for subsequent redundancy analysis. Raw data often exhibit high dimensionality, making redundancy evaluation and mitigation challenging. To address this, domain knowledge-based feature extraction is applied to reduce cross-variable redundancy and dimensionality (e.g., downscaling image or audio). Once dimensionality is reduced, holistic redundancy can be assessed to guide downsampling, removing redundant samples that consume computational resources without contributing to model performance. The downsampled dataset might contain several unbalanced subgroups, which could introduce bias to the ML model. Therefore, this framework mitigates groups-relative redundancy using data synthesis techniques like SMOTE and GAN. Sample-level redundancy must be mitigated before sensor configuration enhancement as data biases can affect cross-sensor redundancy analysis. For example, the importance of a sensor might be overestimated if it primarily captures features pertinent to an overrepresented defect class in the dataset. Sensor configuration enhancement optimizes the monitoring system by adjusting sensor positions and angles or even removing unnecessary sensors based on cross-sensor redundancy analysis, improving efficiency and reducing hardware costs.

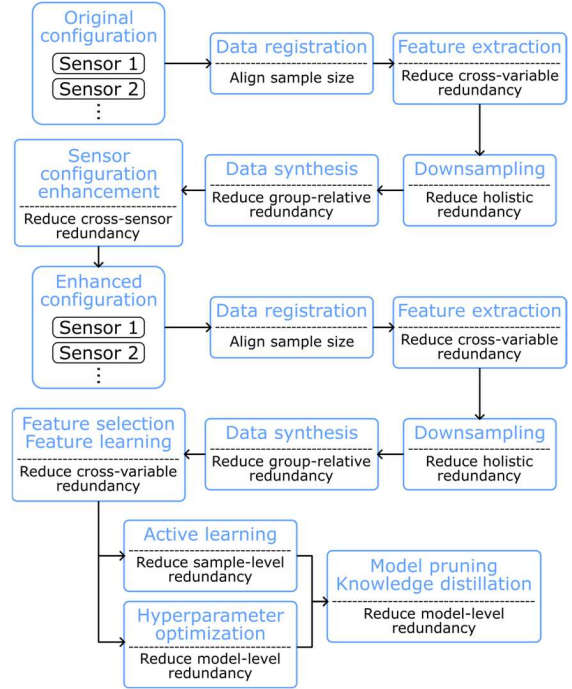

**Figure 3: Multi-level redundancy mitigation framework.**

The data collected from the new sensor configuration undergo the same data preprocessing pipeline—data registration, feature extraction, downsampling, and data synthesis—since sensor configuration enhancement introduces domain shifts and changes the marginal distribution. To further reduce dimensionality and improve efficiency, feature selection and/or feature learning are applied, facilitating active learning and hyperparameter optimization. Active learning acquires new samples with low sample-level redundancy to iteratively and optimally improve model performance using adaptive sampling, which is more effective in a low-dimensional space.

Hyperparameter optimization searches for optimal model configurations, which can incorporate model-level redundancy evaluation and regularization. Lowering cross-variable redundancy enhances the computational efficiency of hyperparameter optimization, potentially reducing the resources required to reach the optimal model configuration. Active learning and hyperparameter optimization are arranged in parallel because both are parts of the training process with no fixed sequence. Afterwards, model pruning or knowledge distillation further reduces model-level redundancy, ultimately providing an efficient final model with optimal size and minimized latency.

The MLRM framework offers a systematic and adaptable pipeline for redundancy analysis and evaluation across all levels. The choice of redundancy evaluation metrics and mitigation techniques should be tailored to the data type, application domain, and specific prediction task. Some steps can be skipped if deemed inapplicable or unnecessary.

## 4. CASE STUDY

This section demonstrates the effectiveness of the MLRM framework using a case study of a multimodal in-situ process monitoring system for DED defect detection. The experiment setup and data acquisition and synchronization are briefly introduced. This section focuses on the identification, evaluation, and mitigation of redundancy.

### 4.1. Experiment setup and data acquisition

This case study leverages an open-access audio-visual dataset collected by the Nanyang Technological University (NTU) during the DED process of 50-layer single-bead walls with maraging steel C300 powder [22]. Pre-optimized process parameters were employed, allowing the natural formation of defects such as cracks and keyhole pores due to heat accumulation. The monitoring system employs two sensors to monitor the real-time melt pool dynamics: 1) a coaxial charge-coupled device (CCD) camera with an acquisition frequency of 30 Hz and 2) a microphone positioned approximately 20 cm from the process zone, operating at a sampling rate of 44,100 Hz. The single-bead walls were wire-cut to reveal their vertical cross-sections, enabling the identification of defect locations across layers along with their spatial coordinates. Optical microscope (OM) images were then captured to detect and localize cracks and keyhole pores within the components, providing physical labels to support ML-based defect detection. The audio-visual signals recorded during the deposition process were synchronized with the robot toolpath coordinates, ensuring precise alignment between the captured data and the deposition trajectory. This synchronization established a direct correlation between observed defects in OM images and the corresponding audio-visual signals recorded along the deposition path.

### 4.2. Multi-level redundancy evaluation and mitigation

Figure 4a shows some examples of the synchronized audio-visual dataset, in which each melt pool image corresponds to an audio snippet of 33.3 milliseconds. A total of 4845 images of 480 × 480 pixels and around 145 seconds of audio were captured during the experiments. Proven efficient for AM defect detection, the audio data were converted to spectrograms, aligning the data types for subsequent analysis [46, 48]. This way, this dataset contains 4845 pairs of melt pool images and spectrograms, in which 1585 pairs are labeled as defect-free samples and 3260 pairs are labeled as defective samples. The dataset was randomly split into a training set, a validation set, and a test set at a ratio of 8:1:1. After data registration, this framework assesses cross-variable redundancy and conducts feature extraction. The high resolution of this dataset is one of the major sources of cross-variable redundancy, resulting in many regions with marginal information (Figure 4a). Thus, the images are downscaled based on the average value within a kernel of 3 × 3 pixels to reduce cross-variable redundancy.

The melt pool images ($X_V$) were downscaled and spectrograms ($X_A$) were created at various sizes (Table 2). Information entropy was utilized to evaluate the randomness of the visual data ($\mathcal{H}(X_V)$) and audio data ($\mathcal{H}(X_A)$) as a measure of information. In this context, downscaling should not significantly change the entropy distribution, characterized by the mean, minimum, and maximum in Table 2, to mitigate redundancy while preserving most information. For both modalities, the mean and maximum entropies were only marginally affected by downscaling, whereas the minimum entropies shifted significantly. Considering the shift of minimum entropy in the visual modality, the redundancy of increasing the image size from 80 × 80 pixels to 480 × 480 pixels is $R_{cv}(X_{V,480}, X_{V,80}) = 1 - \left[\frac{0.294 - 0.245}{0.294}\right] = 0.833$, which indicates a high level of redundancy. For image sizes smaller than 80 × 80 pixels, the visual minimum entropy increases rapidly, suggesting significantly low redundancy and high information loss. For the audio modality, the largest image size is 320 × 320 pixels, beyond which the time interval is too small for creating spectrograms. No significant distribution shift in the entropy is observed when the spectrograms are downscaled to 80 × 80 pixels, making it an appropriate downscaling size. Therefore, both the melt pool images and spectrograms were downscaled to 80 × 80 pixels.

Downsampling and data synthesis were not applied, despite the presence of holistic and group-relative redundancy in the dataset (Figure 4a). Given the relatively small dataset size, downsampling would exacerbate data scarcity. For such a multimodal dataset, it is also challenging to synthesize new samples while preserving the audio-visual correlation. Therefore, we directly conducted sensor configuration enhancement to reduce cross-sensor redundancy. Although sensor positions and settings must remain fixed, it was found that the audio modality could be removed during the operation phase without compromising performance by transferring its knowledge to the visual modality [46]. For instance, utilizing the known correlation between the visual and audio modalities in metal AM process monitoring [13, 14], Xie et al. [46] conducted CMKT to train a model that is robust to noise and solely requires visual input while making predictions.

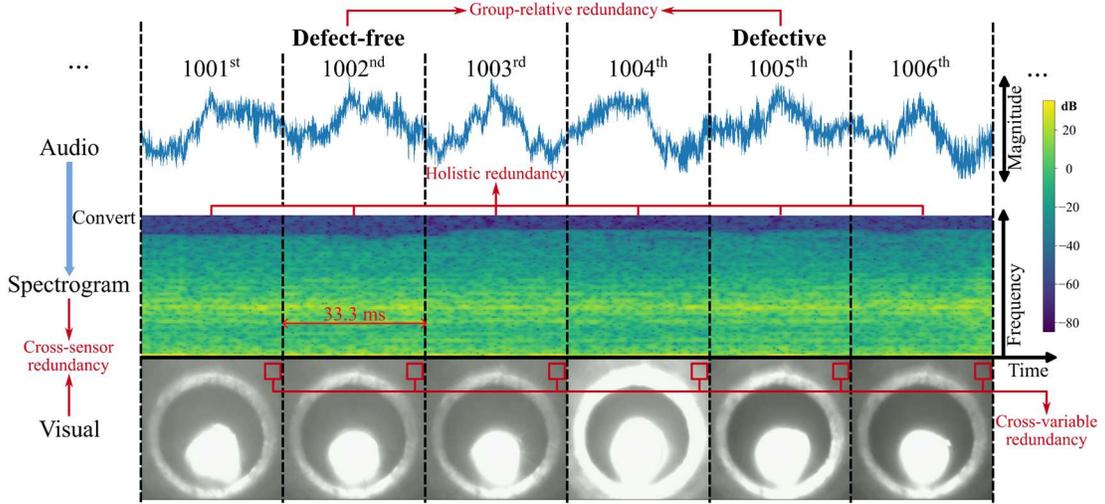

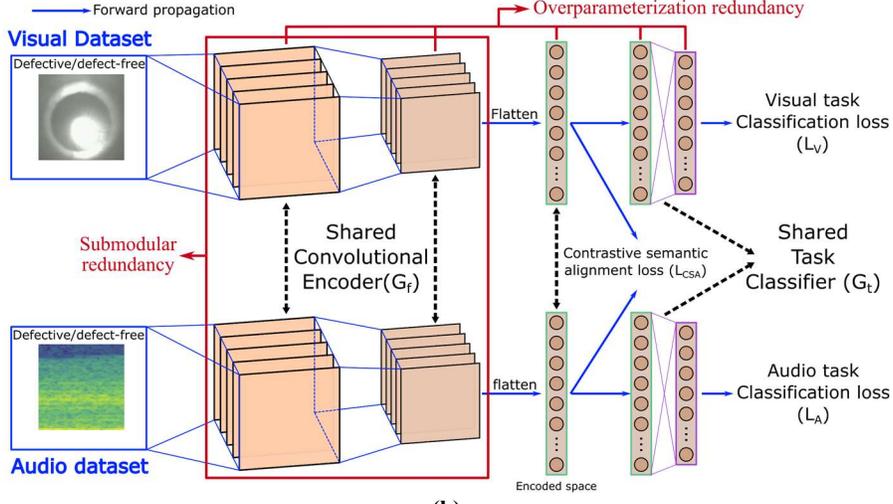

**Figure 4: Schematics of 1) multimodal data alignment; and 2) cross-modality knowledge transfer**

**Table 2: Statistics of the information entropy of the visual and audio data at different image sizes.**

| | Visual entropy (bits) $\mathcal{H}(X_V)$ | | | Audio entropy (bits) $\mathcal{H}(X_A)$ | | |
|---|---|---|---|---|---|---|
| **Image size** | Mean | Min | Max | Mean | Min | Max |
| **20 × 20** | 6.398 | 0.569 | 6.858 | 2.434 | 1.261 | 4.937 |
| **40 × 40** | 6.454 | 0.353 | 6.940 | 2.443 | 1.315 | 4.947 |
| **80 × 80** | 6.416 | 0.294 | 6.946 | 2.538 | 1.637 | 5.210 |
| **160 × 160** | 6.407 | 0.265 | 6.944 | 2.537 | 1.641 | 5.113 |
| **320 × 320** | 6.409 | 0.262 | 6.948 | 2.491 | 1.609 | 5.072 |
| **480 × 480** | 6.391 | 0.245 | 6.930 | NA* | NA | NA |

* NA indicates that the entropy cannot be calculated at this image size.

To mitigate cross-sensor redundancy, we replicated the CMKT method using the same dataset as [46] and provide a brief illustration of the approach in this paper (Figure 4b). A shared convolutional encoder ($G_f$) extracts latent features from both visual and audio inputs and establishes an encoded space for knowledge transfer. Contrastive semantic alignment loss ($L_{CSA}$) was leveraged to align the latent distributions of the visual and audio modalities, encouraging the extraction of 1) modal-shared features that are robust to noise and 2) modal-specific features that effectively separate the defect-free and defective classes. A shared task classifier ($G_t$) utilizes the learned features to classify the input from either modality. As indicated in [46], the CMKT method achieved a balanced accuracy ($\overline{Acc}(\bullet)$) of 98.1%, which was higher than multimodal fusion that requires both modalities for predictions (97.3%). Therefore, the cross-sensor redundancy of adding the audio modality to the system is:

$$R_{cs}(X_A, X_V|Y) = 1 - \left[\frac{(\overline{Acc}(X_V \cup X_A|Y) - \overline{Acc}(X_V|Y))}{|\overline{Acc}(X_V|Y)|}\right] \quad (3)$$
$$= 1 - \left[\frac{0.973 - 0.981}{0.981}\right] = 1.0082 \ .$$

With redundancy larger than one, the audio modality ($X_A$) is fully redundant given the visual modality ($X_V$), and even degrades the model performance regarding the defect detection task ($Y$). Therefore, the microphone is considered redundant during the operation and can be removed without performance compromise using CMKT.

The enhanced sensor configuration trains the model using both modalities but relies solely on the coaxial camera for predictions during the operation phase. Since the marginal distributions of both modalities remained unchanged, the same data registration and feature extraction techniques were applied. Similarly, downsampling and data augmentation were omitted due to the scarcity and heterogeneity of the multimodal dataset. For feature selection and feature learning, CMKT inherently performs feature learning by utilizing an encoder to extract latent features, thereby reducing dimensionality. Active learning was not applied because additional data acquisition was unavailable. Hyperparameter optimization followed the same Bayesian optimization approach as [46], identifying the optimal CMKT model within 300 iterations based on the validation accuracy. However, this optimization process did not incorporate model-level redundancy mitigation.

The CMKT method has effectively eliminated submodular redundancy by employing a shared encoder for both modalities, thereby avoiding separate encoders (Figure 4b). To address the outstanding overparameterization redundancy, model pruning was leveraged to remove close-to-zero parameters in both networks ($G_f$ and $G_t$). The L1 norms of the parameters ($|\theta_1|, |\theta_2|, |\theta_3|, \ldots, |\theta_p|$) were computed and used to increasingly remove the parameters with the smallest L1 norms. Ultimately, it was found that 92.9% and 99.7% of the parameters can be removed from $G_f$ and $G_t$, respectively, without compromising the test balanced accuracy. Therefore, the pruned parameters were considered fully redundant as:

$$R_{op}(\Theta', \Theta_0, Y) = 1 - \left[\frac{\left(\overline{Acc}(\Theta_0|Y) - \overline{Acc}(\Theta_0 \setminus \Theta'|Y)\right)}{|\overline{Acc}(\Theta_0|Y)|}\right] \quad (4)$$
$$= 1 - \left[\frac{0.981 - 0.981}{0.981}\right] = 1,$$

where $\Theta'$ and $\Theta_0$ represent the pruned and original parameters.

Additional to [46], this paper has utilized the MLRM framework to evaluate and mitigate the cross-variable and overparameterization redundancy of the DED process monitoring system. As a result, the computational latency of making predictions for the entire dataset decreased from 68.868 seconds to 6.459 seconds. The storage space required for the entire dataset and the ML model was reduced from 4633.1 MB to 28.58 MB. The detailed breakdown and analysis of the performance improvements were discussed in the next section.

## 5. DISCUSSIONS

Following the proposed MLRM framework, we have systematically addressed multiple forms of redundancy in the audio-visual dataset and ML model. Data registration was applied to align the dataset, downscaling was used to mitigate cross-variable redundancy, CMKT eliminated cross-sensor redundancy, and model pruning effectively reduced overparameterization redundancy. In this section, the performance improvements of the ML-based in-situ DED process monitoring system are demonstrated in terms of the latency, accuracy, and required storage. Figure 5 illustrates the performance changes resulting from the application of specific redundancy mitigation techniques, highlighting their impact on system efficiency. The original system represents the multimodal fusion monitoring system that employs an overparameterized model receiving melt pool image and spectrogram inputs of 320 × 320 pixels.

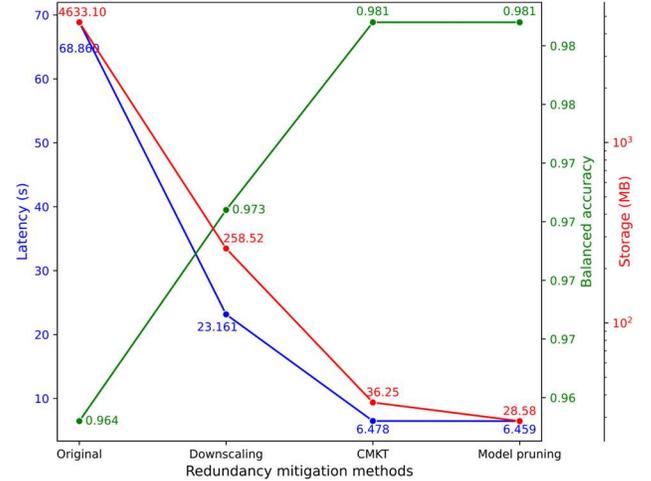

**Figure 5: Performance shifts induced by the MLRM framework.**

The latency of the system is primarily attributed to data preprocessing (e.g., spectrogram conversion, downscaling, and normalization) and model prediction, which were executed on an Intel Core i7 12th Gen processor and an NVIDIA GeForce RTX 4090 GPU, respectively. The latency was measured using the entire dataset, comprising 4845 pairs of melt pool images and spectrograms. Downscaling reduced latency by two-thirds, significantly improving operational efficiency while preserving critical information. CMKT further accelerated the prediction process by eliminating the need for computationally intensive audio-to-spectrogram conversion. Leveraging a high-performance GPU, the total model prediction runtime for the dataset was 0.0460 seconds. Despite pruning 99.4% of the model parameters, the runtime was only reduced to 0.0265 seconds. This is because desktop GPUs are primarily optimized for dense matrix operations, resulting in limited efficiency gains when performing computations on a sparse network after pruning. This framework compressed the defect detection latency per sample from 14.2 milliseconds to 1.33 milliseconds, making it negligible compared with the sampling frequency of the camera (33.3 milliseconds). This improvement is particularly significant for edge computing applications in real-world production environments, where computational resources are limited, and minimizing processing time is essential for maintaining real-time defect detection and process monitoring.

The defect detection accuracy of the system was evaluated based on the balanced accuracy of the test set to account for data imbalance. Downscaling mitigated overfitting by reducing the dimensionality of the input data, leading to an improved test balanced accuracy. CMKT utilized the correlation between the two modalities to extract noise-robust features, further enhancing the balanced accuracy by 0.8%. While reducing the latency and required storage, model pruning did not affect the balanced accuracy of the model. Compared with the original system, the proposed framework decreased the error rate by 47% through redundancy analysis and mitigation, thereby improving the reliability of automated AM quality assessment.

The required storage for deploying and operating the ML-based defect detection model encompasses 1) data storage space for storing input data and 2) memory space occupied by the trained model during execution. Downscaling reduced the dimensionality by 16 times from 320 × 320 pixels to 80 × 80 pixels, substantially decreasing the data storage space needed for the entire dataset by around 18 times. By eliminating the audio modality during operation, CMKT reduced the required data storage space from 226.06 MB to 28.49 MB. The reason why the audio data accounted for a large portion of the storage space was because the spectrograms exhibit considerably higher randomness than the melt pool images. Model pruning further reduced the memory space occupied by the model from 7946 KB to 94 KB, significantly freeing the memory space for other functions of the system. This framework enables the deployment of lightweight quality assessment systems and effectively addresses memory constraints in edge computing environments, making ML-based defect detection more efficient and scalable.

The implications of the MLRM framework extend beyond the abovementioned advantages, influencing various aspects of system deployment and operation (Figure 1). First, the sensor costs were reduced from $10,000 to $5000 by removing the microphone. The labor training and maintenance expenses associated with the audio modality were also eliminated. Additionally, data transmission efficiency was improved substantially due to an 18-fold reduction in data size, leading to lower latency in IoT environments. The combination of a smaller data volume and a lightweight model results in lower energy consumption, making CPSs more resource-efficient. This also contributes to the development of sustainable DTs, optimizing both cost and environmental impact.

## 6. CONCLUSIONS

In this paper, we identified redundancy as a major barrier to the deployment and operation of ML-based process monitoring system in real production environments. However, existing research lacks a unified concept of redundancy and a systematic approach for its analysis and mitigation. To address this gap, we defined redundancy and categorized it into sample-level, feature-level, and model-level redundancy, providing detailed evaluation and mitigation methods for each category. These methods were then integrated into the MLRM framework, a structured approach designed to achieve low-redundancy ML-based process monitoring systems. The effectiveness of the proposed framework was demonstrated through an ML-based in-situ DED defect detection case study, utilizing audio-visual process monitoring data. By applying data registration, downscaling, CMKT, and model pruning, we significantly reduced latency, error rate, and storage requirements. In conclusion, the MLRM framework serves as a fundamental enabler for developing fast, accurate, lightweight, cost-effective, and sustainable ML-based monitoring systems. The limitation of this study is that it focuses on redundancy mitigation, without considering the potential benefits of redundancy. For example, redundancy can be leveraged to quantify uncertainty and enhance system reliability. Future research will explore the value of redundancy, investigating how it can be utilized to further improve system robustness and reliability.

## ACKNOWLEDGEMENTS

Jiarui Xie received funding from Graduate Excellence Award (Grant# 00157) and McGill Engineering Doctoral Award (MEDA) fellowship of the Faculty of Engineering at McGill University. Jiarui Xie also received funding from Mitacs Accelerate Program (Grant# IT13369).

## DECLARATION OF COMPETING INTEREST

The authors declare that they have no known competing interests.

*Corresponding author